\documentclass[aps,pre,reprint,superscriptaddress,showpacs,showkeys]{revtex4-1}

\usepackage{amsmath}
\usepackage[usenames]{color}
\usepackage{amssymb}
\usepackage{graphicx}

\DeclareMathOperator{\e}{e}
\DeclareMathOperator{\W}{W}

\begin{document}

\title{Complete chaotic synchronization and exclusion of mutual Pyragas control\\
in two delay-coupled R\"ossler-type oscillators}

\author{Thomas J\"{u}ngling}
\email{thomas.juengling@physik.uni-wuerzburg.de}
\affiliation{Institute for Theoretical Physics, University of W\"urzburg, Am Hubland, D-97074, W\"urzburg, Germany}
\affiliation{Institute for Solid State Physics, Darmstadt University of Technology, Hochschulstr. 6, D-64289 Darmstadt, Germany}

\author{Hartmut Benner}
\affiliation{Institute for Solid State Physics, Darmstadt University of Technology, Hochschulstr. 6, D-64289 Darmstadt, Germany}

\author{Hiroyuki Shirahama}
\affiliation{Faculty of Education, Ehime University, Bunkyoucho 3, Matsuyama
790-8577, Japan}

\author{Kazuhiro Fukushima}
\affiliation{Faculty of Education, Kumamoto University, Kurokami 2, Kumamoto
860-8555, Japan}

\begin{abstract}
Two identical chaotic oscillators that are mutually coupled via time delayed signals show very complex patterns of completely synchronized dynamics including stationary states and periodic as well as chaotic oscillations.
We have experimentally observed these synchronized states in delay-coupled electronic circuits and have analyzed their  stability by numerical simulations and analytical calculations. We found that the conditions for longitudinal and  transversal stability largely exclude each other and prevent e.g. the synchronization of Pyragas-controlled orbits. Most striking is the observation of complete chaotic synchronization for large delay times, which should not be allowed in the given coupling scheme on the background of the actual paradigm.
\end{abstract}



\maketitle

%
%

\section{Introduction}
\label{introduction}

Chaos control and chaotic synchronization have been one of the most attractive fields of applied nonlinear dynamics for almost two decades in physics \cite{Rosenblum:1997,Hoehne:2007}, engineering \cite{Fukuyama:2002,Fukuyama:2006}, chemical and biological systems \cite{Han:1995,Parmananda:1999} and others. Various strategies aiming at the suppression of undesired chaotic oscillation or at the control of unstable periodic orbits have been developed and have meanwhile been well understood. For a review see \cite{Schoell:2008}. The importance of chaotic synchronization in complex systems has only gradually been clarified \cite{Carroll:1992, Pikovsky:2001,Boccaletti:2002,Yamada:2006}.\\

Chaotic synchronization means that two or more dynamical systems, which are coupled with each other, can correlate their dynamic behavior, even in the case when each system shows irregular behavior. In general, various grades of synchronization can be obtained depending on the coupled elements but also essentially on the form and strength of the couplings. Here we will focus on the strongest form of synchronization called \textit{complete} or \textit{identical} synchronization, which means that the trajectories of the coupled systems coincide exactly. In the case of two identical chaotic systems complete synchronization can be achieved by \textit{non-invasive} couplings. This means that all coupling forces vanish as soon as the synchronized state has been reached, and the synchronized state will be in exact agreement with the previously uncoupled states.\\

Recently the importance of latency effects in the coupling mechanism have attracted major attention both from the aspect of practical applications and for principle reasons. For example, when developing a high speed cryptographic communication system based on chaotic synchronization \cite{Cuomo:1993}, it will be most important to understand the influence of latency times within the whole complex system. Many recent investigations of this problem refer to coupled lasers, e.g. \cite{Heil:2001,Wuensche:2005,Fischer:2006,Klein:2006,Kanter:2007,Kanter:2008,Englert:2010,Englert:2011}, because of their fast characteristic time scales, but also owing to the intrinsic applications of delay lines to make a single laser element chaotic.\\

Delay effects may, however, substantially change the mechanism of chaotic synchronization. It is evident that for two irregular time-dependent signals $\mathbf{x}(t)$ and $\mathbf{y}(t)$ any time delay in the coupling forces has the tendency to create a mismatch in time and will impede the state of complete synchronization. 
This was the reason for the previous opinion that chaotic systems with mutual time-delayed couplings could never synchronize completely. Obviously they cannot synchronize in a non-invasive way, like unstable periodic orbits \cite{Pyragas:1992} or like chaotic systems with instantaneous \cite{Pikovsky:2001} or unidirectional \cite{Voss:2000} couplings. Instead non-zero delay in bidirectional couplings will always result in strong control forces which have to overcome the deviations of identical irregular signals with some time lag. This makes the coupling \textit{invasive}. In this context it has been recently realized that more complicated coupling schemes including additional self-feedback may partly remove the influence of delay from the eqs. of motion of the coupled systems \cite{Juengling:2010}, but even then the principle conflict between delayed couplings and the synchronization of irregular signals is still maintained. Of course there will be further constraints, and we may expect that the tendency of nearby trajectories to exponentially diverge cannot be suppressed by delayed couplings for very large delay times \cite{Kinzel:2009}.\\

In this paper we focus on the most simple coupling scheme and investigate two identical low-dimensional chaotic oscillators that are mutually coupled via a single time-delayed component. We study, both experimentally and analytically, the mutual and auto-synchronization effects that appear in this configuration. Our main interest will be directed to the following problems:
(i) Can such chaotic oscillators completely synchronize at non-zero time delay, and what are the limits for delay time?
(ii) Is it possible to control unstable periodic orbits embedded in these chaotic oscillators and to synchronize them simultaneously?

%
%

\section{Experiment}
\label{roesslers}

\begin{figure}
\begin{center}
\includegraphics[width=0.75\columnwidth]{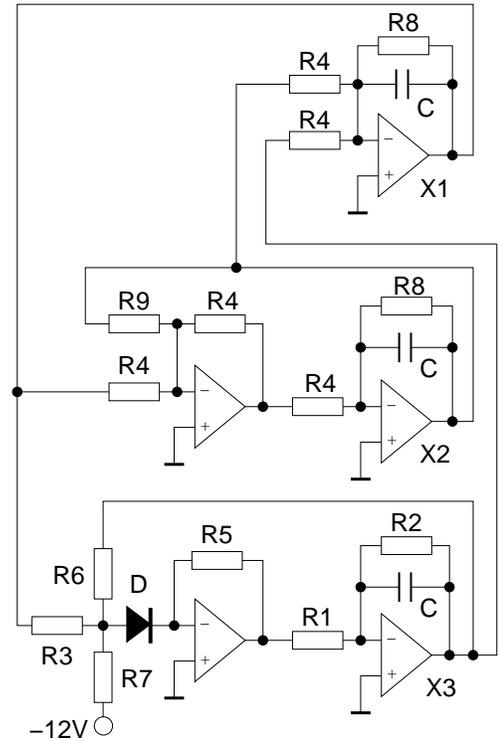}
\end{center}
\caption{Design of a single R\"ossler-type electronic oscillator. Values of components: $R_1=2.7k\Omega, R_2=3.6k\Omega, R_3=7.5k\Omega, R_4=10k\Omega, R_5=13k\Omega, R_6=15k\Omega, R_7=33k\Omega, R_8=200k\Omega$. $R_9$ is variable $0-50k\Omega$ and determines the system parameter $a$ according to $R_9=10k\Omega/a$ (accuracy of resistors 1\%). Capacitators: $C=10nF$ (accuracy 5\%). Type of diode: 1N4007 from DC Components. Type of operational amplifiers: TL084.}
\label{expsch}
\end{figure}

For our experimental investigations we designed two identical autonomous electronic oscillators from analog components including active elements. A diagrammatic view of our setup is shown in fig.~\ref{expsch}.

\begin{figure}
\begin{center}
\includegraphics[width=0.7\columnwidth]{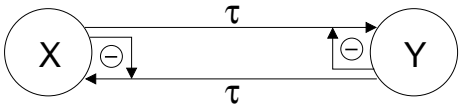}
\end{center}
\caption{Coupling scheme.}
\end{figure}

Following Kirchhoff's rules the equation of motion of this circuit can be written as
\begin{equation}
\begin{split}
\dot{U}_1 & =\omega_0\left( -\alpha \:U_1-U_2-U_3 \right) \\
\dot{U}_2 & =\omega_0\left( U_1+(a-\alpha)\:U_2 \right) \\
\dot{U}_3 & =\beta I_D(U_1,U_3)-\gamma \:U_3 \;.
\end{split}
\label{schwingkreisdgl}
\end{equation}
The typical time scale of chaotic oscillation is determined by the parameter $\omega_0$ which depends on the RC values of integrating op-amps and - for practical reasons - was chosen to be in the order of $1 ms$. The dimensionless parameters $a$ and $\alpha$ are determined by the ratios of resistors. For details see \cite{Juengling:2010}. Here, $a$ is considered as the control parameter of the oscillator. The nonlinear characteristics of the diode D (DC 1N4007) provides the nonlinearity of the circuit. The dynamics of each single oscillator reflects almost perfectly that of a R\"ossler oscillator, in spite of having replaced the multiplicative nonlinearity of the original R\"ossler model by the nonlinear characteristics of a diode. Circuits with such a type of nonlinearity have turned out \cite{Pecora:1997} to be more robust and less sensitive to drifting offsets than circuits including electronic multipliers. Finally, a slight additional damping term $\alpha =0.05$, not present in the original R\"ossler model, has been attributed to the $U_1$ and $U_2$ components in order to avoid a drift in the integrator output signal due to some slight unavoidable offset at the input. The corresponding electronic components of both 'identical' oscillators were compared and carefully selected to guarantee that any deviations remain below 1\%.\\

\begin{figure}
\includegraphics[width=\columnwidth]{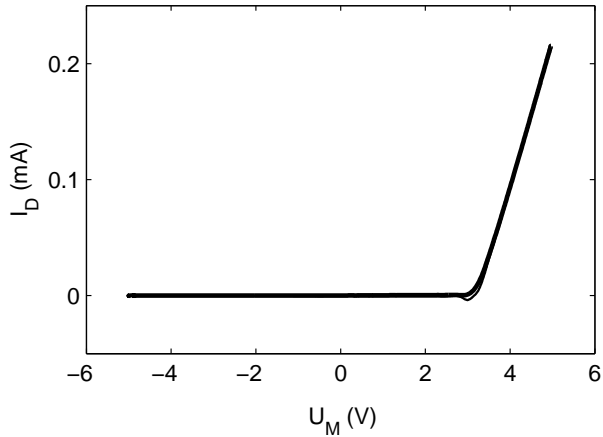}
\caption{Diode characteristics in the network of resistors $R_3,R_6,R_7$. In order to simulate the situation in the running circuit, a sinusoidal test voltage $U_M(t)$ with $5V$ amplitude and a frequency of $1.6kHz$ was applied corresponding to the dynamic input from $U_1$ and $U_3$. At the transition point between locking and conduction ($U_M\approx 3V$) a small undershoot occurs due to the diode capacity, which is negligible for the dynamics of the circuit.}
\label{diodechar}
\end{figure}

The mutual delayed couplings between the oscillators were designed in the simplest possible way via a single component. We chose the $U_2$ components of both oscillators (labeled x and y). The input signals $U_{2,x}$ and $U_{2,y}$ were sent through separate digital delay lines and multiplied by a gain factor $k$ to compose the coupling signals $k\:(U_{2,y}(t-\tau) - U_{2,x}(t))$ and $k\:(U_{2,x}(t-\tau) - U_{2,y}(t))$, respectively. In order to obtain well-defined delay signals the feedback loops generating the control signal are switched on at about one cycle later than the oscillator. Measuring time in units of $\omega_0^{-1}$ and all voltages in volts the coupled system is described by the normalized equation of motion 
\begin{equation}
\begin{split}
\dot{x}_1&=-\alpha x_1-x_2-x_3\\
\dot{x}_2&=x_1+(a-\alpha)\:x_2+k\:(y_{2,\tau}-x_2)\\
\dot{x}_3&=g(x_1,x_3)-c x_3 \\
\\
\dot{y}_1&=-\alpha y_1-y_2-y_3\\
\dot{y}_2&=y_1+(a-\alpha)\:y_2+k\:(x_{2,\tau}-y_2)\\
\dot{y}_3&=g(y_1,y_3)-c y_3 \;.
\end{split}
\label{tworoess}
\end{equation}
Here, $x_i$ and $y_i$ denote the normalized components $U_{i.x}/1V$ and $U_{i.y}/1V$, respectively. The control parameter was fixed to $a=0.265$, $\alpha=0.05$, and $c=\gamma/\omega_0=2.82$. The normalized diode characteristic $g(\cdot)$, which induces the folding process of the R\"ossler-type dynamics, was approximated by a piecewise linear function $g(z_1,z_3)=\beta\left(|z|+z\right)$ with $z=z_1+z_3/2-z_{thr}$.  
$\beta$ and $z_{thr}$ were obtained from a fit of this function to the experimentally obtained profile shown in fig.~\ref{diodechar}. The setting of system parameters, delay time and coupling amplitudes, and the acquisition of data are handled by a separate module, allowing to run the whole experiment automatically.

\section{Observed Phenomena}
\label{overview}

In general there is a large variety of dynamical phenomena arising from this simple setup due to the interplay of nonlinearity and delay. In order to get an overview, we first study the dynamical states occurring on variation of coupling strength and delay time.
\begin{figure}
\includegraphics[width=\columnwidth]{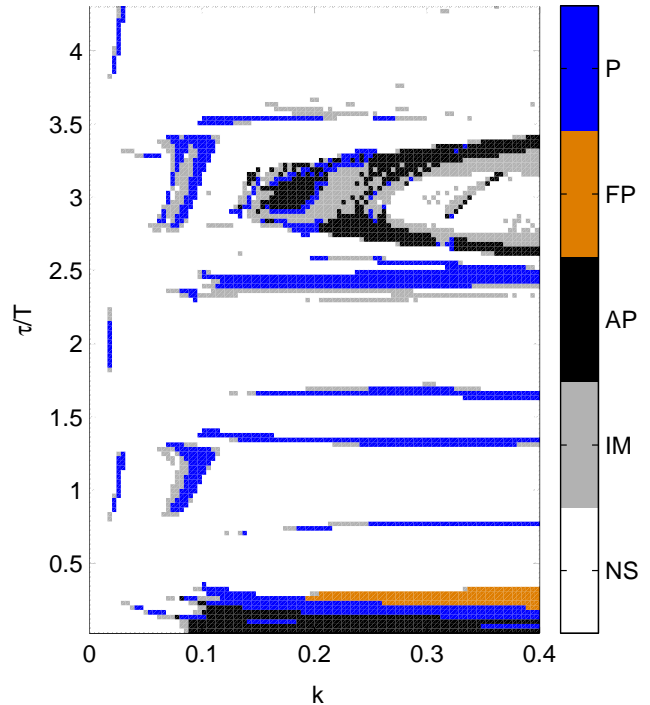}
\caption{(Color online) Experimental diagram of synchronized states. NS: Not completely synchronized, IM: intermittent, AP: aperiodic, FP: fixed point, P: complete periodic synchronization.}
\label{ktauexp1}
\end{figure}
Figure~\ref{ktauexp1} shows an experimental diagram of different dynamical states. Only those parameter combinations are marked for which completely synchronized dynamics is observed, and different colors correspond to different dynamic states, e.g. synchronized orbits of various periodicities as well as synchronized chaos, and - as a trivial limit - also 'synchronized' (i.e. identical) fixed points of both systems. The diagram shows that for the parameter range investigated complete synchronization is rather an exception than the rule.\\
For $\tau=0$, which represents the conventional case without delay, we have complete chaotic synchronization when the coupling strength $k$ exceeds a critical value $k_c$. For delay times slightly larger than zero, we still observe complete synchronization but apart from synchronized chaotic states we also observe synchronized orbits and fixed points. It was counterintuitive to see that periodic orbits do not or almost not synchronize for $\tau=T$ or integer multiples of $T$, but slightly aside there are stripes of complete periodic synchronization. There is also no complete synchronization for half-integer multiples of $T$. The most remarkable phenomenon of CS is a boomerang-shaped domain of complete chaotic synchronization around $\tau=3T$. 
The following sections are devoted to analyze these patterns of complete synchronization.

%
%

\section{Theoretical approach}

In order to discuss the problem in more general terms we start from two identical oscillators $\dot{\mathbf{x}}=\mathbf{f}(\mathbf{x})$ and $\dot{\mathbf{y}}=\mathbf{f}(\mathbf{y})$, which are symmetrically coupled by a bidirectional interaction which is proportional to the instantaneous difference of their trajectories:
\begin{equation}
\begin{split}
\dot{\mathbf{x}}&=\mathbf{f}(\mathbf{x})+\mathbf{K}(\mathbf{y}-\mathbf{x})\\
\dot{\mathbf{y}}&=\mathbf{f}(\mathbf{y})+\mathbf{K}(\mathbf{x}-\mathbf{y})\;.
\end{split}
\label{sch:inst}
\end{equation}
This conventional type of coupling can be considered as the limiting case of zero time delay. The coupling acts as an additional dissipation, and it is well-known that there exist coupling matrices $\mathbf{K}$ such that the completely synchronous state $\mathbf{x}(t)=\mathbf{y}(t)$ becomes stable. If the signal arriving from the other oscillator is delayed in time, e.g. due to finite propagation velocity in a long transmission line, then the ordinary differential eqs.~(\ref{sch:inst}) turn into a system of time-delayed differential eqs.
\begin{equation}
\begin{split}
\dot{\mathbf{x}}&=\mathbf{f}(\mathbf{x})+\mathbf{K}(\mathbf{y}_\tau-\mathbf{x})\\
\dot{\mathbf{y}}&=\mathbf{f}(\mathbf{y})+\mathbf{K}(\mathbf{x}_\tau-\mathbf{y})\;,
\end{split}
\label{sch:hiro}
\end{equation}
where $\mathbf{x}_\tau\equiv \mathbf{x}(t-\tau)$ and $\mathbf{y}_\tau\equiv \mathbf{y}(t-\tau)$. Although such a system is very difficult to tackle, we see immediately that the completely synchronous state $\mathbf{x}(t)=\mathbf{y}(t)$ still represents a solution of eqs.~(\ref{sch:hiro}). Intuitively it is not clear whether  such a state can be stabilized in presence of the delay terms. As we already discussed above, there is a principle difference between the synchronous solution of eq.~(\ref{sch:inst}) and that of eq.~(\ref{sch:hiro}). In the case of instantaneous interactions the coupling forces vanish as soon as the completely synchronized state has been reached, i.e., this kind of coupling is non-invasive. Time-delayed couplings, however, are generally invasive. When complete synchronization has been obtained, the time-delayed signal of the remote system cannot be distinguished from the local signal delayed in time. Thus we may consider the coupling as a \textit{virtual self-feedback}, and the resulting dynamics of the coupled systems eqs.~(\ref{sch:hiro}) represents that of a single delay-system (see e.g.~\cite{Balanov:2005}).\\

Stability analysis of synchronized solutions generally starts from the transformation into appropriate normal coordinates. According to the symmetry of our problem we define the longitudinal coordinates $\mathbf{u}=(\mathbf{x}+\mathbf{y})/2$ and the transversal coordinates $\mathbf{v}=(\mathbf{x}-\mathbf{y})/2$. In terms of theses coordinates the synchronization manifold is determined by $\mathbf{u}=\mathbf{x}=\mathbf{y}$ and $\mathbf{v}=\mathbf{0}$. For small synchronization errors we use the linearized form 
$\mathbf{f}(\mathbf{u}\pm \mathbf{v})=\mathbf{f}(\mathbf{u})\pm \mathbf{Df}(\mathbf{u})\mathbf{v}$, 
so that up to first order in $\mathbf{v}$ the eqs.~(\ref{sch:hiro}) can be written as
\begin{equation}
\begin{split}
\dot{\mathbf{u}}&=\mathbf{f}(\mathbf{u})+\mathbf{K}(\mathbf{u}_\tau-\mathbf{u})\\
\dot{\mathbf{v}}&=\mathbf{Df}(\mathbf{u})\mathbf{v}+\mathbf{K}(-\mathbf{v}_\tau-\mathbf{v})\;.
\end{split}
\label{syncu}
\end{equation}
The equation of motion for $\mathbf{u}$ just reflects that of a single original oscillator with time-delayed self-feedback, which is in exact analogy with the Pyragas control scheme. This form again underlines the - in principle - invasive character of the present coupling due to the additional delay term occurring in the equation. For zero delay, however, the additional control-induced term vanishes and the coupling type becomes non-invasive. The transversal coordinate, $\mathbf{v}$, builds up a special tangent linear system (TLS) driven by $\mathbf{u}$. The stability of its solution $\mathbf{v}=\mathbf{0}$ is determined by the interplay between chaotic drive and coupling-induced damping terms, which include both instantaneous and time-delayed values of $\mathbf{v}$.\\

The stability properties of both eqs. depend in a very complex way on both the delay time $\tau$ and the strength and type of the coupling described by matrix $\mathbf{K}$. The basic criterion for complete synchronization is the stability of the transversal solution $\mathbf{v}=\mathbf{0}$. If the solution of the longitudinal eq. has at least one positive Lyapunov exponent we will observe the synchronization of chaotic trajectories. However, the presence of the self-feedback term may also result in stable periodic solutions
or even fixed points, when the largest Lyapunov exponent becomes zero or negative, respectively. Yet, since both the longitudinal and the transversal eq. depend on the same coupling parameters $\mathbf{K}$ and $\tau$ we have to explore whether the conditions for such regular solutions for $\mathbf{u}(t)$ are consistent with the stability of $\mathbf{v}=\mathbf{0}$. In the following sections we will analyze these different types of solutions separately.

%
%

\section{Stability of the fixed point}
\label{fpoint}

A free running R\"ossler-type system like that of eq.~(\ref{schwingkreisdgl}) has an unstable fixed point close to the origin $\mathbf{u}=\mathbf{0}$. Its exact position depends on the nonlinearity $g(\mathbf{x})$. In our experimental system the fixed point coincides exactly with the origin: $g(\mathbf{0})=0$. We are interested in the longitudinal and transversal stability properties of the synchronized fixed point, i.e. of the solution where the two oscillators lock in their steady states due to the coupling. Note that in the case of steady states our coupling scheme will be non-invasive, since the delay term in the longitudinal part of eqs.~(\ref{syncu}) vanishes. Introducing very small deviations $\delta\mathbf{u}$ and $\delta\mathbf{v}$ from this common fixed point in both the longitudinal and transversal components its stability properties can be obtained through the following linearized equations:
\begin{equation}
\begin{split}
\dot{\delta\mathbf{u}}&=\mathbf{Df}(\mathbf{0})\delta\mathbf{u}+\mathbf{K}(+\delta\mathbf{u}_\tau-\delta\mathbf{u})\\
\dot{\delta\mathbf{v}}&=\mathbf{Df}(\mathbf{0})\delta\mathbf{v}+\mathbf{K}(-\delta\mathbf{v}_\tau-\delta\mathbf{v})\;,
\end{split}
\label{sch:dudv}
\end{equation}
where $\mathbf{Df}(\mathbf{0})$ denotes the Jacobian matrix evaluated at the origin. It creates the common drive for both longitudinal and transversal direction. Due to the simplicity of stationary states the linearized equations  decouple, and we have two independent tangent linear systems that determine the stability of the synchronous fixed point. The Jacobian of our system reads explicitly 
\begin{equation}
\mathbf{Df}(\mathbf{0})=
\begin{pmatrix}
-\alpha & -1 & -1\\
1 & a-\alpha & 0\\
0 & 0 & -c
\end{pmatrix}\;.
\label{fpdrive}
\end{equation}
To further simplify our analysis we neglect the $\delta u_3$ and $\delta v_3$ components, because after short transients both will rapidly relax to zero: $\delta u_3\propto\delta v_3\propto\exp(-c t)$. Only the dominant oscillating part remains relevant for stability and will be treated in terms of a complex-valued normal form:
\begin{equation}
\begin{split}
\dot{z}&=\lambda_0 z+\kappa(+z_\tau-z)\\
\dot{w}&=\lambda_0 w+\kappa(-w_\tau-w)\;,
\end{split}
\label{zw}
\end{equation}
where $z=\delta u_1 +i\delta u_2$ denotes the longitudinal and $w=\delta v_1 +i\delta v_2$ the transversal linearized coordinate. $\lambda_0=(a/2-\alpha)+i$ is one of the complex conjugate eigenvalues of the remaining $2\times2$ Jacobian matrix, and the coupling is $\kappa=k/2$. These results are approximations of the real dynamics for the case of strong mixing of the components due to rotation. An exact transformation is not possible because $\mathbf{Df}$ and $\mathbf{K}$ do not commute. Our approximation is valid as long as the coupling strength is not too large. The corresponding eigenvalues of eqs.~(\ref{zw}) are determined by
\begin{equation}
\begin{split}
\lambda_\parallel&=\lambda_0-\kappa(1-e^{-\lambda_\parallel\tau})\\
\lambda_\bot&=\lambda_0-\kappa(1+e^{-\lambda_\bot\tau})\;.
\end{split}
\label{zwev}
\end{equation}

\begin{figure}
\includegraphics[width=\columnwidth]{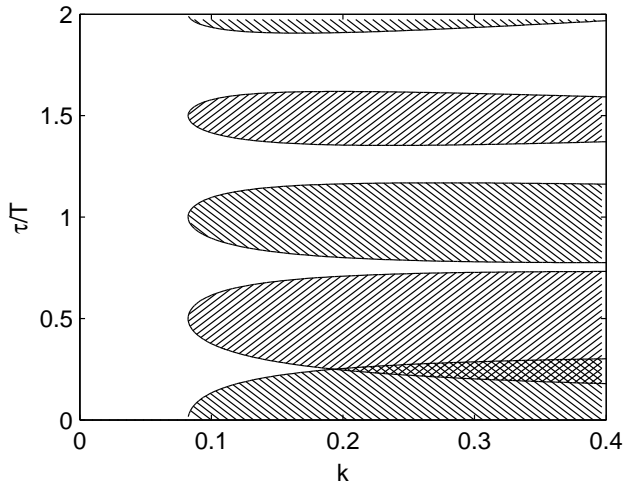}
\caption{Longitudinal (///) and transversal ($\backslash\backslash\backslash$) stability domains of the fixed point.}
\label{fpuv}
\end{figure}

\begin{figure}
\includegraphics[width=\columnwidth]{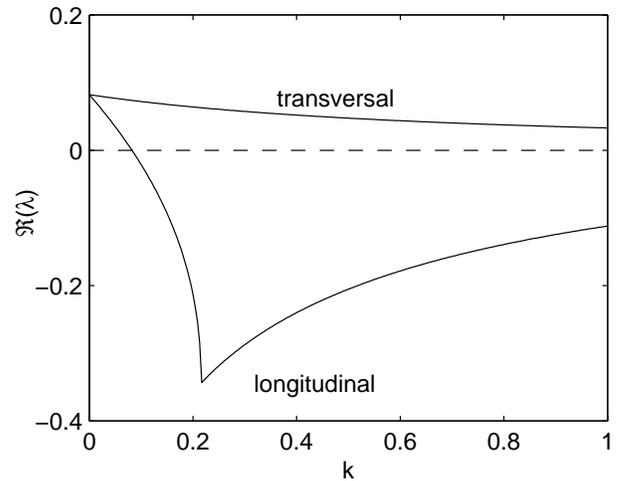}
\caption{Real part of the largest eigenvalue $\Re(\lambda)$ from eq.~(\ref{zw}) for both longitudinal and transversal subspace at a half-period delay $\tau=T/2$.}
\label{ltl05}
\end{figure}

The synchronized fixed point is stable if all $\Re(\lambda)<0$. Figure~\ref{fpuv} shows the stability domains in the $k,\tau$-plane. The domains of transversal stability are centered at integer multiples of $T$, while the domains of longitudinal stability are centered at half-integer multiples. There is only a very small overlap between both kinds of  domains at a quarter-period delay $\tau=T/4$. In fig.~\ref{ltl05} we have illustrated the $k$-dependence of both the longitudinal and transversal Lyapunov exponents at a half-period delay $\tau=T/2$. For the longitudinal component we recognize the profile familiar from the Pyragas control of a periodic orbit with $\pi$-torsion \cite{Just:1997}. But for the transversal component the delay term is of opposite sign, so it counteracts with the instantaneous term - just like in the case of torsion-free orbits, which cannot be stabilized by ordinary Pyragas control.

The overlap at the quarter-period delay arises from a $\pi/2$ phase rotation, for which the opposite signs in eqs.~(\ref{zwev}) lead to complex conjugate eigenvalues. We found that such an overlap can only occur as long as $\Re(\lambda_0)\tau\le0.326$ \cite{Juengling:2010}. This explains the missing overlaps at larger delay times. Altogether, the experimental observation of synchronous fixed points in a very small regime at $\tau \approx T/4$ only is well understood in terms of our stability analysis.

%
%

\section{Periodic synchronization}
\label{persync}

Each chaotic oscillator has included an infinite set of unstable periodic orbits. In this section we focus on the low-periodic orbits, which are embedded in the chaotic attractor of an uncoupled oscillator. There exists at least one period-1, period-2, etc. orbit with a period, which is close to an integer multiple of the mean cycle time $\bar{T}\approx 2\pi/\omega_0$. Moreover, since in our circuits these unstable orbits arise from a period-doubling route to chaos, they all have a finite torsion, which means that their neighborhood undergoes a $\pi$-flip during each cycle $T_n$. Therefore these orbits should be accessible to Pyragas control under the common restrictions. In fact, when we applied Pyragas control with $\tau=T_n$ and appropriate coupling strength to a single oscillator, the stabilization of such orbits could be observed in experiment. But when trying to synchronize two identical Pyragas-controlled orbits by means of our mutual delay-coupling scheme eq.~(\ref{tworoess}) - which should be  non-invasive in this case, too - we always failed. This becomes evident from fig.~\ref{ktauexp1} for $\tau=T$, $2T$, and $4T$. Note that the narrow synchronization windows observed at smaller $k$-values merely refer to delay-induced periodic solutions for which the feedback force does not vanish. So we conclude that for our coupling scheme the non-invasive synchronization of stabilized periodic orbits is not possible. This can be explained by analogy with the synchronization mechanism of fixed points.
Let $\boldsymbol{\xi}(t)$ denote the unstable periodic orbit of an uncoupled oscillator with $\boldsymbol{\xi}(t+T)=\boldsymbol{\xi}(t)$. Then for $\tau=nT$, $n$ integer, the trajectory $\mathbf{x}(t)=\mathbf{y}(t)=\boldsymbol{\xi}(t)$ is a solution of eqs.~(\ref{sch:hiro}) which we want to analyze. Introducing small longitudinal and transversal deviations $\delta\mathbf{u}$ and $\delta\mathbf{v}$ from the orbit, we arrive at almost the same two tangent linear systems as in eqs.~(\ref{sch:dudv}) but with a time-periodic drive instead of a constant one.
\begin{equation}
\begin{split}
\dot{\delta\mathbf{u}}&=\mathbf{Df}(\boldsymbol{\xi}(t))\delta\mathbf{u}+\mathbf{K}(+\delta\mathbf{u}_\tau-\delta\mathbf{u})\\
\dot{\delta\mathbf{v}}&=\mathbf{Df}(\boldsymbol{\xi}(t))\delta\mathbf{v}+\mathbf{K}(-\delta\mathbf{v}_\tau-\delta\mathbf{v})\;.
\end{split}
\label{xidudv}
\end{equation}
The longitudinal equation represents exactly the Floquet problem well-known for Pyragas control. So within the synchronization manifold the trajectory shows the same stability properties as an uncoupled oscillator with self-feedback. In particular, considering only the longitudinal stabilization we have the limitation $\Re(\lambda_0)\tau\le2$. For details we refer to ref.~\cite{Just:1999}. However, there is still the transversal solution to be stabilized. The transversal equation coincides almost completely with the longitudinal one, but the delay term has the opposite sign. This means for our Floquet problem that, whenever the $\pi$-torsion of a periodic orbit provides successful stabilization in the longitudinal direction, the inverted delay term and the torsion cancel each other in the transversal subspace, just as if the orbit was torsion-free. The stability dilemma is exactly the same as for the fixed point of our system at half-period delay: For the presented coupling scheme at least one direction of the complete phase space remains unstable. But this is not an absolute restriction. Stable synchronized periodic oscillations may occur when the delay time is detuned from the orbit period, as can be seen in fig.~\ref{ktauexp1}. As mentioned in the preceding section, there could be an overlap between the domains of longitudinal and transversal stability, but only if the restriction  $\Re(\lambda_0)T\le0.326$ is fulfilled together with a $\pi/2$ torsion. For fixed points the overlapping domains were located around a quarter-period delay. For periodic orbits the situation is more complicated, since the orbits also get deformed by the invasive feedback for $\tau\neq n \cdot T$. But as a rough estimate the $\pi/2$ argument holds for the stability domains of the period-1 and period-2 orbits as well.

%
%

\section{Complete chaotic synchronization}
\label{hirowindow}

Complete synchronization of chaotic trajectories is observed for very small delay times $\tau\approx0$ but also for rather large delay times $\tau\approx3T$, see fig.~\ref{ktauexp1}. The first result is intuitively expected since it occurs close to the limit of conventional chaotic synchronization. This process is well established, and a slight change of the synchronization signal should not immediately spoil the whole mechanism. The synchronization at $\tau\approx3T$, however, was extremely surprising, even more when taking into consideration that the delay time exceeds the 'chaotic memory', i.e. the Lyapunov time, of the uncoupled chaotic oscillator, by about factor 2. We evaluated this characteristic time to amount to $1.6 T$ \cite{Juengling:2010}. So it appears impossible that the tendency of nearby trajectories to exponentially diverge could be suppressed by coupling signals for such a large delay time \cite{Kinzel:2009}. What is also obvious for time-delayed chaotic synchronization, we do have invasive couplings. In fact, we do not synchronize the two original oscillators but - according to eq.~(\ref{syncu}) - two delay systems, which for increasing delay time and coupling strength will increasingly differ from the original oscillators. 

In order to analyze the observed synchronization phenomenon more in detail we again consider the tangent linear systems
\begin{equation}
\begin{split}
\dot{\delta\mathbf{u}}&=\mathbf{Df}(\boldsymbol{\zeta}(t))\delta\mathbf{u}+\mathbf{K}(+\delta\mathbf{u}_\tau-\delta\mathbf{u})\\
\dot{\delta\mathbf{v}}&=\mathbf{Df}(\boldsymbol{\zeta}(t))\delta\mathbf{v}+\mathbf{K}(-\delta\mathbf{v}_\tau-\delta\mathbf{v})\;,
\end{split}
\label{uvtls}
\end{equation}
which only differ from the previous forms, eqs.~(\ref{sch:dudv}) and eqs.~(\ref{xidudv}), by the common chaotic drive $\boldsymbol{\zeta}(t)$. The trajectory $\boldsymbol{\zeta}(t)$ obeys the equation of motion
\begin{equation}
\dot{\boldsymbol{\zeta}}(t)=\mathbf{f}\left(\boldsymbol{\zeta}(t)\right)+\mathbf{K}\left(\boldsymbol{\zeta}(t-\tau)-\boldsymbol{\zeta}(t)\right)\;,
\label{zetaeq}
\end{equation}
which is the longitudinal part of eqs.~(\ref{syncu}) and includes the effect of the coupling parameters. Note that $\boldsymbol{\zeta}(t-\tau)\neq\boldsymbol{\zeta}(t)$ for all $\tau\neq0$, which excludes any non-invasive delayed couplings. The solutions $\delta\mathbf{u}(t)$ and $\delta\mathbf{v}(t)$ reflect the stability properties of $\boldsymbol{\zeta}(t)$. All we can say is that there is some general tendency for $\delta\mathbf{u}(t)$ and $\delta\mathbf{v}(t)$ to diverge (or converge) exponentially with the largest Lyapunov exponents $\Lambda_\parallel$ and $\Lambda_\bot$ of the longitudinal and of the transversal subspace, respectively. A necessary criterion for chaotic synchronization is $\Lambda_\bot<0$ while $\Lambda_\parallel>0$. The challenging question is, can we present any limitations in terms of $\Lambda_\parallel$ and $\Lambda_\bot$ for the domains of complete chaotic synchronization, as we could give for synchronized fixed points or periodic orbits? Certainly not in a strict way. Since the TLS, eqs.~(\ref{uvtls}), are not driven by constant or periodic terms, which implies that $\delta\mathbf{u}_\tau\neq\mathbf{A}\cdot\delta\mathbf{u}$ and $\delta\mathbf{v}_\tau\neq\mathbf{A}\cdot\delta\mathbf{v}$ for any constant matrix $\mathbf{A}$, we cannot directly derive characteristic equations like eqs.~(\ref{zwev}). 
Yet, in the appendix, we develop an approximate form for a similar type of equations, which will allow us to specify limiting conditions for the domains of complete chaotic synchronization.

\begin{figure}
\includegraphics[width=\columnwidth]{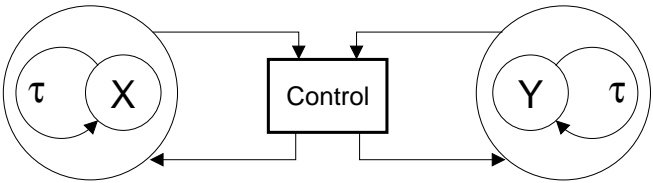}
\caption{Gedankenexperiment: Two delay systems coupled via a non-invasive controller.}
\label{fig:gedexp}
\end{figure}

A first guess of such a limiting condition can be obtained by means of a simple gedankenexperiment (see fig.~\ref{fig:gedexp}): Imagine a setup, which consists of two identical oscillators like those used in our experiment. Let each of them include a self-feedback with fixed values of time delay $\tau$ and coupling $\mathbf{K}$ identical for both circuits. We consider this feedback to be part of each system, i.e. we have two identical delay systems, each of them characterized by their common largest Lyapunov exponent $\Lambda_\parallel$. When $\Lambda_\parallel$ is positive, both delay systems are chaotic and, starting from neighboring initial conditions, they separate exponentially with this rate. The divergence of the trajectories can be compensated by an additional active or passive controller that applies an appropriate dissipative force to each oscillator. E.g. by means of a non-invasive control force $\mathbf{K}'(\mathbf{y}-\mathbf{x})$ proportional to the difference between their instantaneous signals complete synchronization can be achieved, and the delay systems show the same dynamics as if they were not influenced by the controller. Now, let us assume that the controller has also a finite latency $\tau'$ between measuring $\mathbf{x}$ and $\mathbf{y}$ and feeding the control forces back to the systems. Then the corresponding equations of motion take the form
\begin{equation}
\begin{split}
\dot{\mathbf{x}}&=\mathbf{F_K}(\mathbf{x},\mathbf{x}_\tau)+\mathbf{K}'(\mathbf{y}_{\tau'}-\mathbf{x}_{\tau'})\\
\dot{\mathbf{y}}&=\mathbf{F_K}(\mathbf{y},\mathbf{y}_\tau)+\mathbf{K}'(\mathbf{x}_{\tau'}-\mathbf{y}_{\tau'})\;.
\end{split}
\label{controllereqs}
\end{equation}
where $\mathbf{F_K}(\mathbf{x},\mathbf{x}_\tau)$ and $\mathbf{F_K}(\mathbf{y},\mathbf{y}_\tau)$ denote the delay systems. It is a well-known fact in control engineering (see e.g.~\cite{Gu:2003} and references) that this latency $\tau'$ has to remain below some critical value in order to successfully stabilize the completely synchronized state $\mathbf{x}=\mathbf{y}$ - i.e., returning to our previous notation, to stabilize the fixed point $\mathbf{v}=0$. If the instability were created by an ordinary system without delay one would expect that
\begin{equation}
\lambda\;\tau'<1
\label{lt1}
\end{equation}
must be satisfied, where $\lambda$ is the real part of the corresponding exponent of the unstable fixed point the controller has to stabilize. By identifying the instability of the fixed point $\mathbf{v}=0$ with the (longitudinal) Lyapunov exponent of our delay systems, $\lambda\equiv\Lambda_\parallel$, we obtain an estimate for the largest delay time $\tau'$ accessible by non-invasive control. This analytical result is quite general and not restricted to the specific model of our experiment.

If $-$ like in our virtual setup $-$ $\lambda$ itself is largely determined by the intrinsic time delay $\tau$, which is of the same order as $\tau'$, then we expect an interplay between the two different sources of delay, which might  modify the conventional limitation (\ref{lt1}). In particular, for $\tau'=\tau$ and $\mathbf{K}'=\mathbf{K}$ the eqs.~(\ref{controllereqs})  just represent our original coupling scheme, eq.~(\ref{sch:hiro}). A more sophisticated approach presented in the appendix results in a moderate extension of the simple estimate:
\begin{equation}
\Lambda_\parallel\;\tau<1.28\;,
\label{lpt13}
\end{equation}
which means that the interplay between the non-invasive controller delay $\tau'$ and the intrinsic delay $\tau$ results in a 30\% increase of the stability limit.\\

\begin{figure}
\includegraphics[width=\columnwidth]{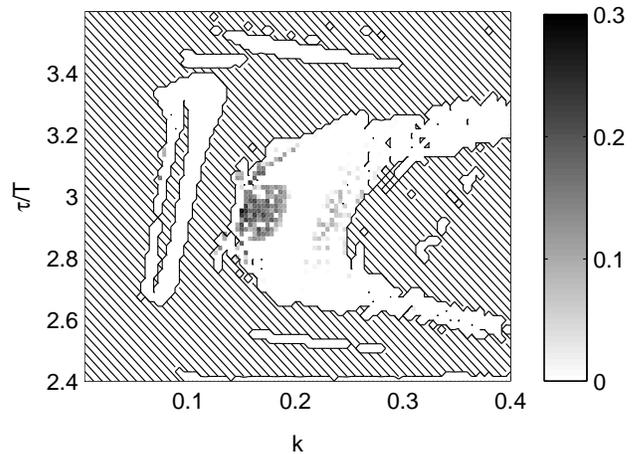}
\caption{Numerical simulation of the product $\Lambda_\parallel\cdot\tau$ shown in grayscale for synchronized solutions at $\tau\approx3T$. Hatched areas indicate absence of complete synchronization.}
\label{hslt1}
\end{figure}

The idea of coupled delay systems provides the key for understanding the possibility of chaotic synchronization at large delay times, e.g. at $\tau\approx3T$. Self-feedback in our delay systems creates a general tendency to reduce the grade of chaos and to make the systems more laminar. This is no surprise since the self-feedback acts just like the mechanism of Pyragas control. We have studied this reduction in detail by calculating numerical solutions $\boldsymbol{\zeta}(t)$ of eq.~(\ref{zetaeq}) and analyzing their longitudinal stability according to eq.~(\ref{uvtls}). Results are presented in fig.~\ref{hslt1} in grayscale. In order to limit our efforts we have only considered such sets of coupling parameters that result in complete synchronization.    

The synchronization patterns obtained by numerical simulations coincide almost exactly with our experimental data, which means that the numerical model yields a quite realistic description of the experiment. Within the synchronization domains relation~(\ref{lpt13}) is clearly satisfied. For most of the parameter values, where an aperiodic synchronous signal was detected, the exponent is close to zero, so the trajectories are either weakly chaotic or quasiperiodic. The occurrence of quasiperiodicity might be related to the dynamics observed at the upper edge of a Pyragas control regime, where a controlled orbit becomes unstable through a Hopf bifurcation and the dynamic switches over to a torus. At those parameter sets, where the completely synchronized state remains clearly chaotic, the Lyapunov exponent $\Lambda_\parallel(k,\tau)$ of this feedback-disturbed chaos is still markedly smaller than the exponent of the original undisturbed oscillator, $\Lambda_0$. From our simulations $\Lambda_0$ was evaluated to amount to $0.1 \omega_0$, which corresponds to a Lyapunov time $T_{\Lambda_0}\equiv 1/\Lambda_0 = 1.6\:T$. This means that for the $k-\tau$ area investigated the 'chaotic memory' of the delay system has been extended in time by more than factor 5, which explains the possibility that complete chaotic synchronization can, in principle, occur for such large delay times 
$\tau\approx3T$ and for even larger ones. 

\begin{figure}
\includegraphics[width=\columnwidth]{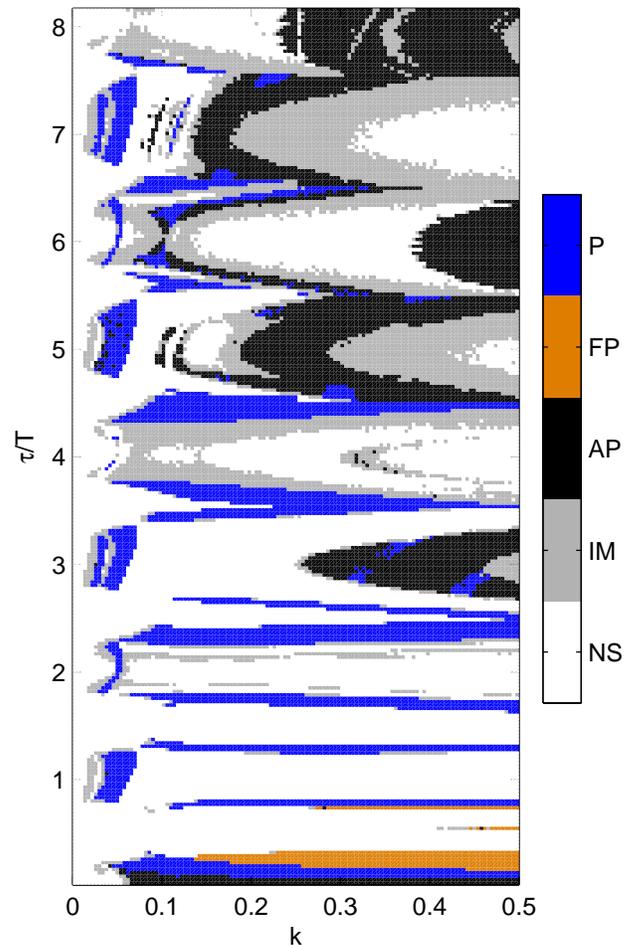}
\caption{(Color online) Dynamical states for system parameter $a=0.22$. Labels as in fig.~\ref{ktauexp1}.}
\label{fig:a022}
\end{figure}

Finally we underline that complete chaotic synchronization, as observed for $\tau\approx3T$, is not a singular event. For larger delay times we may observe even more extended domains of complete synchronization - provided that the constraint~(\ref{lpt13}) is still satisfied. In fig.~\ref{fig:a022} we present additional experimental data taken in the same setup, but for a different system parameter, $a=0.22$. For this value of $a$ the uncoupled oscillators are still unstable, but less chaotic than for our reference system ($a=0.265$). The figure shows a lot of parameter domains in $k$ and $\tau$ where aperiodic synchronization was observed. In our automatic scans of the $k-\tau$ plane we did not systematically distinguish between quasiperiodic and chaotic time series, but among those selected time series, which were analyzed in detail, we found many examples of complete chaotic synchronization. Complete aperiodic synchronization is again observed in a boomerang-shaped domain at $\tau\approx3T$, but also in regions at higher multiples of $T$, where the shape of the domains differ qualitatively for odd and even $T$. The increased size of the synchronization domains is obviously related to the weaker chaotic behavior of the uncoupled oscillators. Because of that the resulting longitudinal Lyapunov exponents $\Lambda_\parallel$ should be still smaller than for our reference system and - according to condition~(\ref{lpt13}) - synchronization should be accessible for even larger delay times.     

%
%

\section{Conclusions}

Two chaotic oscillators that are mutually coupled via time delayed signals show very complex patterns of completely synchronized dynamics, including fixed points, periodic orbits and quasi-periodic as well as chaotic trajectories. We have experimentally observed these synchronization phenomena in two delay-coupled R\"ossler-type electronic circuits and analyzed the synchronized states by means of analytical calculations and numerical simulations of the corresponding stability equations. We found that the equations determining the longitudinal and the transversal stability only differ by the sign of a single delay term, which, however, in the case of fixed points and periodic orbits results in limitations, which contradict or widely exclude each other. For instance, it is impossible to completely synchronize two Pyragas-controlled orbits, because the torsion, which favorites longitudinal stability, will prevent transversal stabilization and vice versa. For fixed points a similar phase argument excludes the simultaneous stabilization of longitudinal and transversal components for all but a very small set of coupling parameters.\\

For chaotic trajectories such a dilemma does not exist, since merely transversal stability has to be achieved. Even more, if one of the stability exponents is allowed to be positive, this phase argument provides for a better chance to find the other one negative. Therefore, complete chaotic synchronization should occur more frequently, as long as the delay time does not exceed the limit of the chaotic instability. In fact, we observed complete synchronization of chaotic trajectories in our experiment for delay times far beyond the timescale of the uncoupled oscillators. This unexpected behavior could be attributed to the invasive character of our coupling scheme in the case of aperiodic trajectories. The coupling forces act like a virtual self-feedback, which has the tendency to reduce the grade of chaotic instability and to increase the Lyapunov time. As a consequence complete chaotic synchronization could be achieved for very large delay times. We derived a general limitation for $\tau$ connecting the transversal stability with the longitudinal Lyapunov exponent of the coupled chaotic systems.\\
 
In spite of the simple and highly symmetric coupling scheme complete synchronization remains a rather exceptional phenomenon, which shows a very complex dependence on the coupling parameters. The complexity seems to be characteristic for systems with time delay and might result from the interplay of diverging tendencies: to complicate the system by inflating the dimension of phase space, or to simplify the system through the action of an invasive self-feedback. We were able to present general arguments excluding or limiting complete synchronization for different types of dynamics. However, the detailed mechanism of transversal stabilization is not yet fully understood and has certainly to include more detailed information on the phase rotation (see \cite{Pyragas:2008}) and chaotic mixing of the components. Further investigations based on a stochastic modeling of the chaotic drive might shed more light on this problem.

%
%

\section*{Appendix}
\label{app:ltaulimit}

Starting from the tangent linear systems, eqs.~(\ref{uvtls}), we consider a small variation of the coupling strength $k$ without changing the driving trajectory $\boldsymbol{\zeta}(t)$, which is determined by eq.~(\ref{zetaeq}) for some fixed value of $k$. Such a  virtual step can be implemented only numerically, because in experiment - due to the invasive character of the coupling - any variation of $k$ would always influence $\boldsymbol{\zeta}(t)$. By analogy with the procedure for periodic orbits in~\cite{Just:1997} we introduce an approach for the Lyapunov exponents in the sense of a linear expansion in $k$ around $k=0$
\begin{equation}
\Lambda=\Lambda_0+k\left(\chi\e^{-\Lambda\tau}-\eta\right)\;.
\label{lambdaeq}
\end{equation}
Here $\Lambda_0$ depends only on the trajectory and is considered as a hidden intrinsic property. If $\boldsymbol{\zeta}$ were an unstable \textit{orbit} of a single oscillator, $\Lambda_0$ would correspond to the real part of its Floquet exponent in the absence of control. $\chi,\eta\in\mathbb{R}$ are expansion coefficients quantifying the  \textit{coupling efficiencies} of time-delayed and instantaneous terms, respectively. In general we have $\chi\neq\eta$, which reflects the fact that for an aperiodic trajectory the delayed coupling term cannot be mapped to the instantaneous term, as mentioned above. Note that in contrast to the corresponding Floquet problem \cite{Just:1997} eq.~(\ref{lambdaeq}) holds for Lyapunov exponents and - provided that $k$ remains sufficiently small - all involved parameters are real. In terms of such an approach we estimate the limits of transversal stability for a given longitudinal instability $\Lambda_\parallel$ and delay time $\tau$. We start from eq.~(\ref{lambdaeq}), which holds for both longitudinal and transversal dynamics
\begin{equation}
\begin{split}
\Lambda_{\parallel}&=\Lambda_0+k\left(\chi\e^{-\Lambda_\parallel\tau}-\eta\right)\\
\Lambda_{\bot}&=\Lambda_0+k\left(\gamma\chi\e^{-\Lambda_\bot\tau}-\eta\right)\;.
\end{split}
\label{ltgamma}
\end{equation}
Note that the symmetry of the longitudinal and transversal LTS results in identical expansion parameters $\chi$ and $\eta$. The coefficient $\gamma$ includes a variation of sign and strength of the transversal coupling with respect to the longitudinal coupling and can also be applied to more general schemes than considered here \cite{Englert:2011}.  
The instantaneous terms in the coupling scheme can be easily included in the intrinsic Lyapunov exponent through the renormalization  
$\Lambda_0 \rightarrow \Lambda_0 -\eta k :=\Lambda_0'$. 
Explicit solutions of these transcendental eqs. can be given in terms of the Lambert W-function:  
\begin{equation}
\begin{split}
\Lambda_\parallel&=\Lambda_0'+\frac{1}{\tau}\W\left(k\tau\e^{-\Lambda_0'\tau}\right)\\
\Lambda_\bot&=\Lambda_0'+\frac{1}{\tau}\W\left(\gamma k\tau\e^{-\Lambda_0'\tau}\right)\;.
\end{split}
\end{equation}
The possibility of complete synchronization depends on the fact whether the global minimum of $\Lambda_\bot$ can become negative. For given $\Lambda_0'$ the minimum is determined by the properties of the W-function and occurs at $W(-\e^{-1})=-1$, i.e., if $k$ takes such a value that $\gamma k\tau\e^{-\Lambda_0'\tau}=-1/e$ we have 
\begin{equation}
\Lambda_{\bot,min} = \Lambda_0'-\frac{1}{\tau} \;,
\end{equation}
and correspondingly,
\begin{equation}
\Lambda_\parallel = \Lambda_0'+\frac{1}{\tau}\W\left(\frac{-1}{e\gamma}\right) \;.
\end{equation}
Eliminating $\Lambda_0'$ from both eqs. we finally arrive at the relation
\begin{equation}
\Lambda_{\bot,min}=\Lambda_\parallel-\frac{1}{\tau}\left[1+\W\left(\frac{-1}{\e\gamma}\right)\right] \;,
\end{equation}
which connects the minimum value of $\Lambda_\bot$ with the corresponding longitudinal exponent $\Lambda_\parallel$ and the delay time $\tau$. Stable transversal dynamics only occurs for $\Lambda_{\bot,min}<0$, which directly leads to a general limitation for the delay time:
\begin{equation}
\Lambda_\parallel\cdot\tau < 1 + \W\left(\frac{-1}{e\gamma}\right) \;.
\end{equation}
In our present coupling scheme we have $\gamma = -1$, which results in the limiting condition  
$\Lambda_\parallel\cdot\tau < 1+\W(+1/e) \approx 1.28$.

%
%


\end{document}